\documentclass[final]{elsarticle}

\usepackage{lineno}
\usepackage[hidelinks]{hyperref}

\journal{XXX}

\usepackage{algorithmic}
\usepackage{algorithm}
\usepackage{color}
\usepackage{multirow}
\usepackage{makecell}
\usepackage{verbatim} 
\usepackage{subcaption}
\usepackage[export]{adjustbox}
\usepackage{booktabs}
\usepackage{amssymb}
\usepackage{bbm}
\usepackage{float}
\usepackage{graphicx}
\usepackage{amsmath}

\newcommand{\etal}{\textit{et al}. }

\usepackage[normalem]{ulem}

\newcommand{\ADD}[1]{#1}

\begin{document}

%
%
%
%
%
%
%

\begin{frontmatter}

%
%
%
%
%
%
%
%
%

\title{Semi-supervised cross-lingual speech emotion recognition}

\author[aff1]{Mirko Agarla}
\ead{m.agarla@campus.unimib.it}
\author[aff1]{Simone Bianco}
\ead{simone.bianco@unimib.it}
\author[aff1]{Luigi Celona\corref{cor1}}
\ead{luigi.celona@unimib.it}
\author[aff1]{Paolo Napoletano}
\ead{paolo.napoletano@unimib.it}
\author[aff2]{Alexey Petrovsky}
\ead{alexey@petrovsky.eu}
\author[aff1]{Flavio Piccoli}
\ead{flavio.piccoli@unimib.it}
\author[aff1]{Raimondo Schettini}
\ead{raimondo.schettini@unimib.it}
\author[aff3]{Ivan Shanin}
\ead{ivan.shanin@gmail.com}

\address[aff1]{University of Milano - Bicocca, Milano, Italy}
\address[aff2]{Independent researcher}
\address[aff3]{Institute of Informatics Problems, FRC CSC RAS}
\cortext[cor1]{Corresponding author.}

\newpageafter{author}
\begin{abstract}
Performance in Speech Emotion Recognition (SER) on a single language has increased greatly in the last few years thanks to the use of deep learning techniques.
However, cross-lingual SER remains a challenge in real-world applications due to two main factors: the first is the big gap among the source and the target domain distributions; the second factor is the major availability of unlabeled utterances in contrast to the labeled ones for the new language. Taking into account previous aspects, we propose a Semi-Supervised Learning (SSL) method for cross-lingual emotion recognition when only few labeled examples in the target domain (i.e. the new language) are available. Our method is based on a Transformer and it adapts to the new domain by exploiting a pseudo-labeling strategy on the unlabeled utterances. In particular, the use of a hard and soft pseudo-labels approach is investigated. We thoroughly evaluate the performance of the proposed method in a speaker-independent setup on both the source and the new language and show its robustness across five languages belonging to different linguistic strains. The experimental findings indicate that the unweighted accuracy is increased by an average of 40\% compared to state-of-the-art methods.
\end{abstract}

\begin{keyword}
semi-supervised domain adaptation, speech emotion recognition, cross-lingual, semi-supervised learning
\end{keyword}

\end{frontmatter}


\section{Introduction}
SER is a fundamental aspect of computational paralinguistics as it concerns the analysis of the non-verbal elements of speech \citep{schuller2013computational}. SER, which aims to infer the emotional state of a speaker \citep{el2011survey}, could support a wide range of domains, including human-computer interaction \citep{schuller2013computational}, healthcare \citep{tumanova2020effects}, and public safety \citep{lefter2017aggression}. For instance, SER systems could be employed in interactive dialogue systems to make them empathetic \citep{bertero2016real}, in healthcare systems for the diagnosis of disorders and diseases \citep{hansen2022generalizable}, and in commercial applications for detecting customer satisfaction in call-centers and by employment agencies to find suitable candidates \citep{perez2021user}. Existing SER models have achieved satisfactory results for valence/arousal estimation \citep{xiao2016speech} and emotion classification \citep{scheidwasser2022serab} when the training and test data are from the same corpus. However, the performance of these models degrades when applied to new corpora of same/different languages due to domain shift \citep{feraru2015cross}. This problem occurs mainly in real-world scenarios where the people using a given SER system may differ or speak languages other than those used to train the system.
 
 Over the years, several methodologies have been developed to speed up the adaptation of a pre-trained system to new people or a new language by leveraging semi-supervised/incremental learning~\citep{zhang2016enhanced} and transfer learning~\citep{feraru2015cross}. Numerous approaches have been proposed to reduce the domain shift problem for cross-corpus or cross-lingual SER, namely eliminate or reduce the difference between the source and target data distribution~\citep{cai2021unsupervised,tamulevivcius2020study}. Most of these approaches are based on deep learning techniques as they generally prove to be more effective than traditional machine learning techniques also for SER~\citep{tamulevivcius2020study}. Supervised Domain Adaptation (SDA) methods for SER exploit labeled utterances of the target corpus to adapt the recognition model to work properly on the new set of data~\citep{neumann2018cross,zhou2019transferable,tamulevivcius2020study}. However, these methods require the new language utterances to be labeled, which may not be possible as their collection is expensive. Therefore, a more practical solution is Unsupervised Domain Adaptation (UDA) which only demands unlabeled utterances from the new language. Many UDA methods try to reduce the distribution shift between the source and target languages~\citep{latif2019unsupervised,cai2021unsupervised,li2021unsupervised,ocquaye2021cross}.

\begin{figure*}
    \centering
        \begin{subfigure}{.4\textwidth}
            \centering
            \includegraphics[width=\columnwidth]{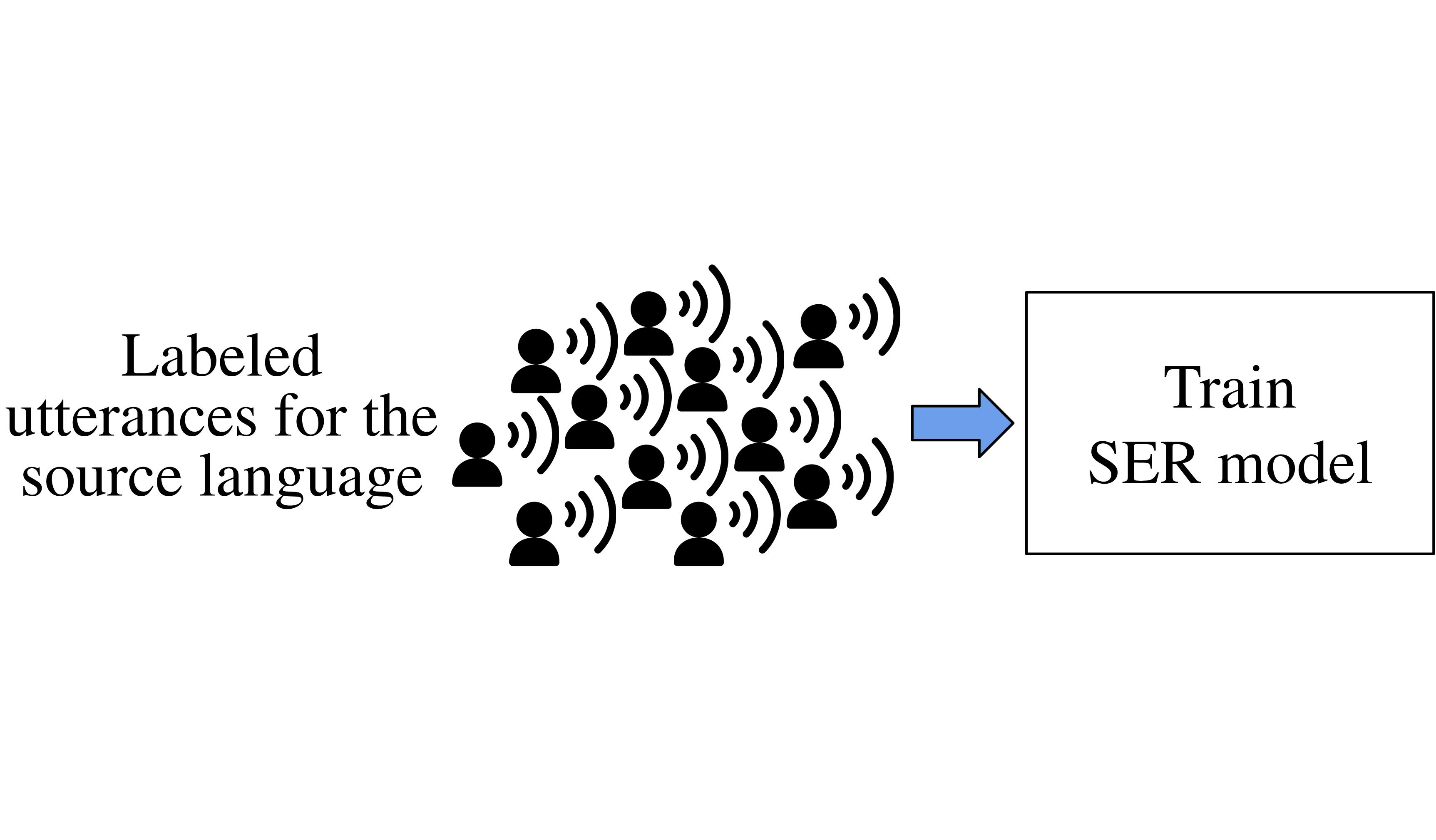}
            \caption{Single-language SER}
            \label{fig:single-language}
        \end{subfigure} \hspace{3.5em}
        \begin{subfigure}{.4\textwidth}
            \centering
            \includegraphics[width=\columnwidth]{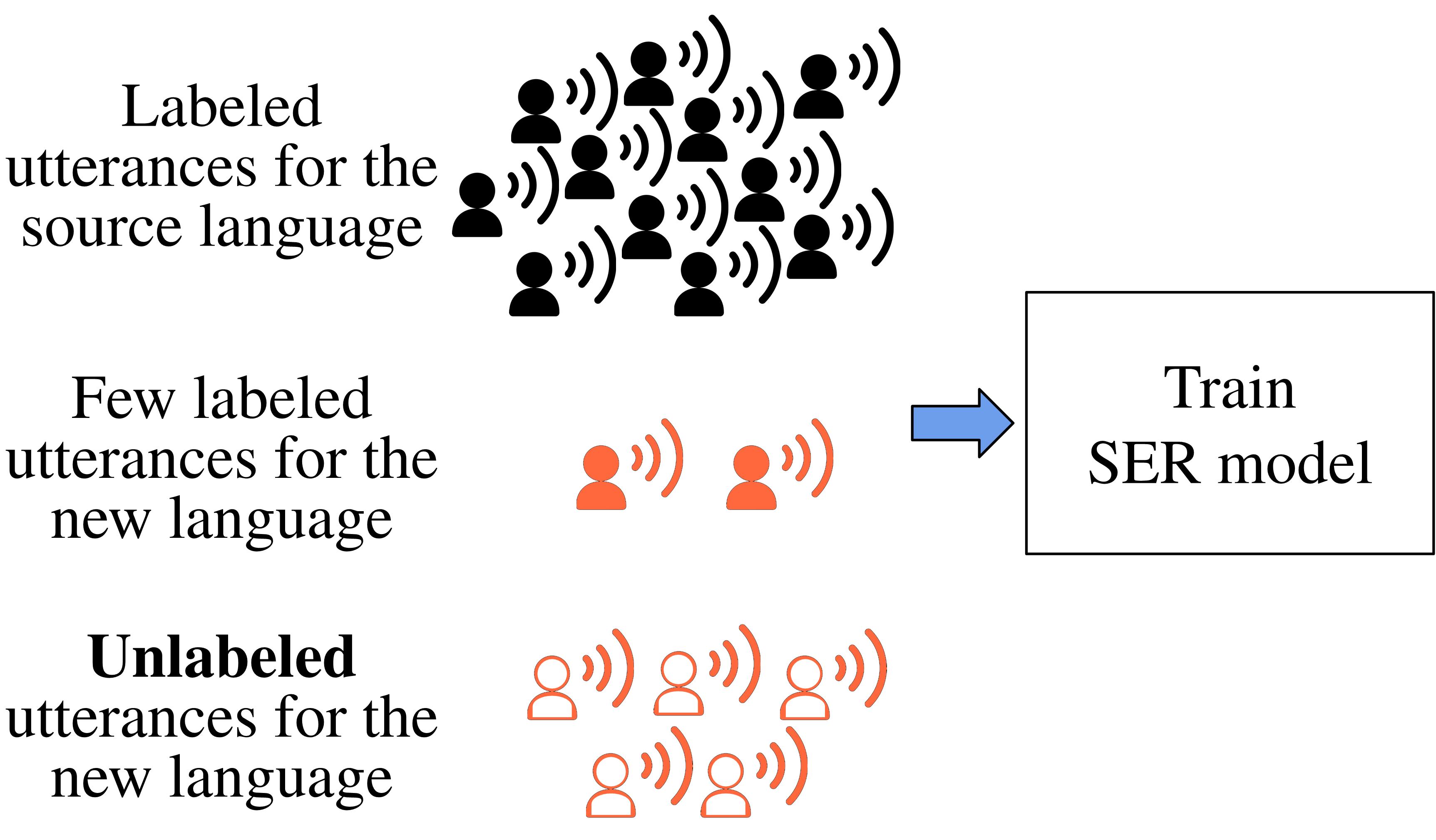}
            \caption{Train on a new language}
            \label{fig:new-language}
        \end{subfigure}
    \caption{Pipeline of the proposed cross-lingual SER method. (a) The model is first trained on the source language and (b) it is then adapted to the new language for which a few labeled utterances are available.}
    \label{fig:teaser}
\end{figure*}

In this paper we formulate the cross-lingual SER as a SSL problem. This scenario assumes that for the new language there are few labeled and many unlabeled utterances. We first train a deep learning based SER model on the source language dataset in which all utterances are annotated with the emotion label (see Fig. \ref{fig:single-language}). The SER model is then adapted to a new language for which the emotion of most training utterances is unknown. The labeled data of the first language are available (see Fig. \ref{fig:new-language}). Pseudo-labeling is adopted to generate labels for the unlabeled utterances and guide the learning process. Unlike most cross-lingual SER methods which focus on the binary classification of valence, our approach deals with the prediction of five emotion categories. In our experiments we consider English as the source language since it is the most widespread language in the world \citep{english2021}.

The proposed method for cross-lingual SER based on pseudo-labeling is suitable for use in all-day consumer technologies, such as smartphones, smartmirrors, and smartwatches. These devices collect massive amounts of unlabeled data, making traditional supervised learning methods difficult to implement. The proposed method overcomes this challenge by requiring only small amounts of labeled data and large amounts of unlabeled data, lowering manual annotation costs and shortening data preparation time. As a result, the method can be used in consumer technologies at a much lower cost than traditional supervised learning-based methods, making it a more practical and accessible solution. By implementing this method, consumer technologies can accurately recognize and respond to emotions expressed in different languages, improving communication and user experience. For example, a smartmirror with cross-lingual SER could provide personalized recommendations based on a user's emotional state, or could adjust lighting and temperature to create a more comfortable environment based on the user's emotions.

Apart from a method for cross-lingual SER, this work provides an analysis of different models as utterance encoder. In particular, it is demonstrated that a Transformer-based utterance encoder trained to build meaningful representations of speech boosts the performance compared to state of the art methods. Furthermore, in the adaptation procedure it is verified that balancing pseudo-labeled vs. labeled utterances helps to improve the generalization capabilities of the learned model.

To summarize, the main contributions of this paper are:
\begin{itemize}
    \item A cross-lingual SER framework spanning five languages.
    \item A SSL based cross-lingual SER method for emotion categorization.
    \item The experimentation of several utterance encoders, i.e. a CNN for speech emotion classification, a CNN and a Transformer trained for speech representation learning.
    \item Two different approaches for generating pseudo-labels are investigated.
    \item An utterance rebalancing strategy to reduce the cardinality gap between the labeled utterances available for the source language and the labeled or pseudo-labeled utterances for the new language.
    \item A thorough analysis of how the variation in the number of labeled utterances for the new language impacts performance.
\end{itemize}
The rest of the paper is organized as follows. Section \ref{sec:related-work} introduces some previous works on cross-lingual emotion recognition. In Section \ref{sec:method}, SSL based cross-lingual SER is formalized and then the proposed method is described. Experimental setup and result analysis are presented in Sections \ref{sec:experiments} and \ref{sec:results}, respectively. Finally, we conclude in Section \ref{sec:conclusions}.
\section{Related work}
\label{sec:related-work}
%

Cross-lingual SER methods involve the use of two languages, the source language for which the emotion information is available for all the samples and the target language, for which only few labeled samples are available. The aim of cross-lingual SER is then to learn from the source language and extend the learned knowledge on the unlabeled samples of the target language. This propagation is an active process that involves the use of the few samples available and, if present, of auxiliary information available for both languages. At each learning episode, unlabeled samples of the target language which are believed to belong to an emotional class, are labeled. This process is called pseudo-labeling and in turn will support future learning episodes for evaluating the remaining samples. The choice of the acoustic features is quite important and must be kept into account \citep{tamulevivcius2020study}.

The primary taxonomy of cross-lingual SER methods is given by the strategy used to convey knowledge on the new language. Two are the main schemes used in the state of the art. The first one is the use of auxiliary information that is available on both languages to learn a shared feature space. In this context, the emotion recognition task and the side task are performed simultaneously through a joint-training.
Cai \etal~\citep{cai2021unsupervised} proposed a neural network with two branches, leading respectively to emotion and language classification. During training, the first branch is trained only with the source language corpus with emotion labels while the second by both languages. This training schema allows to exploit all existing information of the first and the second language to create a shared feature space. Gradient Reversal Layer (GRL) is adopted to force the features to be meaningful for the primary task (emotion classification) and at the same time to be indistinguishable for the auxiliary task (language classification). Performance is measured in terms of valence and arousal on Urdu, Estonian, IEMOCAP, Persian, and German (4 emotions).
Li \etal~\citep{li2021unsupervised} proposed an SSDA memory-based system called Neural Network with Pseudo Multilabel (NNPM). In first place, they use a siamese network with self attention for projecting source and target utterances in a learned feature space. Then, the source-domain features are dynamically stored in the dynamic external memory. Emotion similarity is gained through cosine distance between features in the memory and of the target utterance. Pseudo-labels are given to the target domain utterances based on the similarity score. Hard negative sample mining strategy is used to improve the learning whereas the features result less representative. Performance is measured with weighted and unweighted accuracy.
Ocquaye \etal \citep{ocquaye2021cross} exploited joint training to perform SSDA. A neural network with one common branch and a set of task-specific branches is proposed. Two branches perform emotion recognition respectively on the source samples and the pseudo-labeled target samples. The evaluation has been conducted on SAVEE, IEMOCAP, EMO‐DB, FAU‐AIBO (German), and EMOVO for valence classification.

The second strategy widely used in the context of cross-lingual SER is the use of adversarial training. This technology showed great ability in domain transfer and thus is very effective in this scenario. 
Latif \etal~\citep{latif2019unsupervised} proposed a method for learning a language-independent emotion recognition feature vector in the context of UDA. This system is based on Generative Adversarial Networks (GANs) as this technology has shown great potential in learning the underlying data distribution. Specifically, the proposed method has two generators to project respectively the source and the target utterance in a common feature space that is later evaluated through a critic. The feature space is then constrained to carry emotion information through a classification loss. In other words, the adversarial loss makes the feature space homogeneous both for the source and the target languages while the classification loss makes the features meaningful for the task of emotion recognition. Performance assessed on EMOVO, SAVEE, Urdu, and EMO-DB for valence classification. 

Domain Adversarial Neural Network (DANN) is a method that generates domain invariant feature representations. This allows to reduce the gap among source and target domain features \citep{abdelwahab2018domain}.

However, the effectiveness of domain adversarial training strongly depends on the distribution of the two databases: in fact, adversarial attacks and instabilities may occur in the training phase if the data points are significantly different from each other. Aggregate multi-task Learning (AL) is another technique that has been used to improve the generalization of the trained model by incorporating information of gender and naturalness \citep{kim2017towards}.

Extending the work of Sung \etal~\citep{sung2018learning}, Ahn \etal~\citep{ahn2021cross} presented Few-shot Learning and Unsupervised Domain Adaptation (FLUDA), which aims to train an embedding and a metric module that respectively project the utterances in a meaningful shared feature space and learn the differences among classes. The embedding and metric module are optimized to predict class similarity for each episode by exploiting few samples composing the support set and pseudo-labels assigned in the previous episode. During training, an auxiliary module is used to determine whether the labeled sample is real or pseudo-labeled. The proposed method estimates four categorical emotions (neutral, happy, sad and angry) and uses IEMOCAP and CREMA-D as source corpora while MSP-IMPROV, EMO-DB or KME were used as target corpus. 
However, the samples in few-shot learning significantly depend on the choice of the support set, that can make its application challenging to a practical setup. Furthermore, the strong assumption that the support set is uniformly sampled from a single distribution, leads to the selection of an unstable number of samples for each class during  training. With the aim of solving the previous challenges, Zhou \etal~\citep{zhou2019transferable} used adversarial network to perform SSDA. Specifically, a GAN is modified such that the generator projects the utterance in a feature space carrying emotion information and the critic has to determine which emotion class belongs the input feature and the used language. This stage is trained with the source language for which the labels are known. In second place the critic is frozen and the encoder is adapted to the new language. This second step forces the encoder to adapt and generate compatible features with the ones obtained in the first step. The method has been benchmarked on EMO-DB and Aibo in terms of positive and negative emotions.

Recently, Das \etal~\citep{das2022towards} presented a Variational AutoEncoder (VAE) for learning a latent space able to discriminate emotions and to generalize on different languages simultaneously. They achieved this goal by (i) exploiting a Kullback-Leibler (KL) loss annealing using cyclic scheduling to improve emotion discrimination, (ii) employing semi-supervised training of the VAE by incorporating a clustering loss in the learning function. Experimental results have been collected for IEMOCAP, SAVEE, EMO-DB, CaFE, and AESD in terms of four emotions. Kshirsagar \etal~\citep{kshirsagar2022cross} explored the combined use of Bag-of-Words (BoW) methodology, domain adaptation and data augmentation as strategies to counter the damaging effects of cross-lingual SER. The authors also proposed a new method called N-CORAL in which all languages are mapped to a common distribution. Experiments with the German, Hungarian, Chinese, and French languages show the advantages of the proposed N-CORAL method, combined with data augmentation and BoW for valence-arousal estimation.

The related works can be thus summarized as follows:
\begin{itemize}
    \item Several studies in the literature demonstrate that it is possible to perform cross-language SER without labeled utterances for the new language using auxiliary information, generative methods, or adversarial and few-shot learning.
    \item Domain adaptation approaches based on adversarial neural networks are widely used for cross-lingual SER; however, there is still room for performance improvement.
    \item The use of the newer Transformer architectures for utterance encoding has not been explored and used for cross-lingual SER.
\end{itemize}
\section{Method}
\label{sec:method}
The problem of cross-lingual SER is first formulated in Section \ref{sec:problem-formulation} and then the proposed SSL method for cross-lingual SER is described in Section \ref{sec:method-description}.

\subsection{Problem formulation}
\label{sec:problem-formulation}
We represent sets with special Latin characters (e.g., $\mathcal{S}$). Lower or uppercase normal fonts, e.g., $K$ denote scalars. Matrices are in uppercase bold letters (e.g., $\mathbf{M}$), while lowercase bold letters represent vectors as in $\mathbf{v}$. We use lowercase Latin letters to represent indices (e.g., $i$).

We formulate our cross-lingual SER as the following domain adaptation task. We have a source language corpus with $N_s$ labeled utterances as source domain, $\mathcal{D}_s = \{(\mathbf{X}^s_i, y^s_i)\}_{i=1}^{N_s}$, and a new language corpus, $\mathcal{D}_t$, as target domain. The new language corpus, $\mathcal{D}_t = \{\mathcal{U}_t \cup \mathcal{K}_t\}$, consists of a set $N_u$ of unlabeled utterances $\mathcal{U}_t = \{\mathbf{X}^t_i\}^{N_u}_{i=1}$, and a set $N_k$ of labeled utterances $\mathcal{K}_t = \{\mathbf{X}^t_i,y^t_i\}^{N_k}_{i=1}$. The number of utterances of the source language $N_s$ is much higher than the number of labeled utterances for the new language $N_k$, i.e. $N_s \gg N_k$. Utterances $\mathbf{X}^s_i$ and $\mathbf{X}^t_i$ are elements of $\mathbb{R}^{F \times T}$, where $F$ and $T$ are the number of frequency bins and the number of time frames, respectively. The utterance labels $y^s_i$ and $y^t_i$ are scalar values such that $y \in \mathbb{Z}:1\le y \le C$ where $C$ is the number of emotion categories within the corpora. We consider that the source and target corpora contain the same number $C$ of emotion categories. 

Our goal is to learn a reliable emotion classifier on $\mathcal{D}_s$, $\mathcal{U}_t$, and $\mathcal{K}_t$, which preserves performance on source language $\mathcal{D}_s$ and generalizes well on $\mathcal{D}_t$.

\subsection{SSL for cross-lingual SER}
\label{sec:method-description}
The proposed semi-supervised cross-lingual SER is a deep learning model $f$ parameterized with $\theta$ that maps an input utterance $\mathbf{X}$ into a basic emotion $y$, $y=f(\mathbf{X}, \theta)$.

\begin{figure*}
    \centering
    \includegraphics[width=\linewidth]{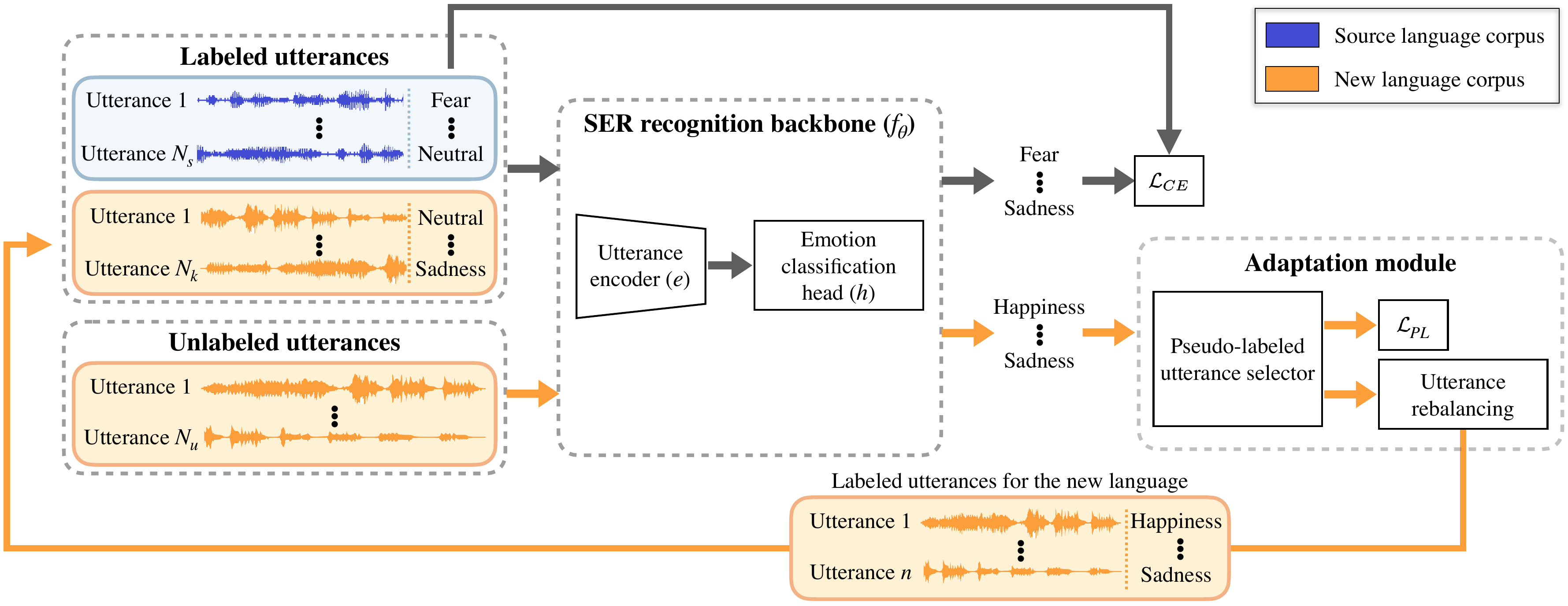}
    \caption{The pipeline of our method.}
    \label{fig:method}
\end{figure*}

As depicted in Figure \ref{fig:method}, our method consists of two modules, namely the \textit{SER recognition backbone} and the \textit{adaptation module}. The SER recognition backbone ($f_\theta$) classifies emotions for both the source and new language utterances. During the training phase, we introduce the adaptation module to improve the discriminative power and generalization ability of $f$ on the new language. The adaptation module relies on a pseudo-labeling strategy to allow model training on the unlabeled utterances of the new language. Furthermore, we include an utterance rebalancing mechanism to avoid that the model is biased on the source language due to the higher cardinality of utterances compared to those of the new language.

In the next sections we detail the previously introduced modules.
\subsubsection{SER recognition backbone}
It is the core of our method that deals with utterance classification. It consists of two modules, i.e., the utterance encoder $e$ and the classification head $h$. The utterance encoder $e$ is a deep architecture like a CNN or a Transformer that takes a raw waveform or an audio representation as input, $\mathbf{X}$, and returns a $d$-dimensional feature vector $\mathbf{x} \in \mathbb{R}^D$. The choice of the utterance encoder is important for the performance of the proposed method. Therefore, three different deep architectures are considered, namely EmotionCNN \citep{tamulevivcius2020study}, Bootstrapping Your Own Latent for Speech (BYOL-S) \citep{scheidwasser2022serab}, and Hidden-unit BERT (HuBERT) \citep{hsu2021hubert}. \textbf{EmotionCNN} is a CNN architecture for cross-lingual speech emotion recognition. It is composed of three convolutional layers. Each convolutional layer is followed by a ReLU, a batch normalization layer, and $3\times3$ max pooling, respectively. The model is fed with a cochleagram-based representation of the raw waveform and outputs a 128-dimensional feature vector. \textbf{BYOL-S} is a CNN model for audio representation inspired by the Bootstrapping Your Own Latent (BYOL) model initially proposed for self-supervised image classification~\citep{grill2020bootstrap}.
It is trained on the speech utterances of the AudioSet \citep{gemmeke2017audio} dataset.
BYOL-S is currently the state of the art in SER \citep{scheidwasser2022serab}. The model accepts input utterances of variable length and returns a single 1024-dimensional feature vector per input utterance. All utterances are converted to a log-scaled Mel spectrogram with a window size of 64ms, hop size of 10ms, and mel-spaced frequency bins $F = 64$ in the range 60–7,800 Hz. Each spectrogram is normalized by subtracting the mean and dividing by the estimated standard deviation for the frames of the spectrogram. \textbf{HuBERT} is a Transformer-based approach for self-supervised speech representation learning. 
It consists of a convolutional waveform encoder, a BERT-Base encoder \citep{devlin2018bert}, a projection layer, and a code embedding layer. It is trained on the Librispeech 960h dataset \citep{panayotov2015librispeech} to classify randomly masked frames to pseudo-labels. The labels are generated by running K-Means clustering with 100 clusters on 39-dimensional MFCC features. The model accepts raw waveforms of variable length as input and returns a single 768-dimensional feature vector per input utterance. The main characteristics of the three architectures are summarized in Table \ref{tab:encoder-comparison}.
\begin{table}
    \centering
    \caption{Utterance encoder architectures considered within the proposed method.}
    \label{tab:encoder-comparison}
    \resizebox{\columnwidth}{!}{
    \begin{tabular}{lccc}
    \toprule
         &  \makecell{EmotionCNN\\ \citep{tamulevivcius2020study}} & \makecell{BYOL-S\\ \citep{scheidwasser2022serab}} & \makecell{HuBERT\\ \citep{hsu2021hubert}} \\ \midrule
         Input & Cochleagrams & MFCC & Waveform \\
         Architecture & CNN & CNN & Transformer \\
         Feature vector dim. & 128 & 1024 & 768 \\
         Pretraining & -- & AudioSet & Librispeech 960h \\
         ML paradigm & -- & self-supervised & self-supervised \\
         Num. of params & 35,584 & 1.6M & 95M \\
         \bottomrule
    \end{tabular}}
\end{table}

The feature vector obtained from one of the previously described utterance encoders is processed by the classification head $h$ to predict $\mathbf{p} \in \mathbb{R}^C$, i.e., the probability distribution over the $C$ emotion categories. The classification head consists of a linear layer followed by a softmax:
\begin{equation}
    \mathbf{p} = \mathrm{Softmax}(h(\mathbf{x}, \theta_h)),
\end{equation}
where $\theta_h$ is the set of weights $\mathbf{W} \in \mathbb{R}^{D \times C}$ and bias $\mathbf{b} \in \mathbb{R}^C$.

\subsubsection{Adaptation module}
This module aims to exploit the unlabeled utterances of the new language in the model training. To this purpose, an SSL approach is used to generate pseudo-labels $\tilde{y}$ for the unlabeled utterances of the new language dataset, $\mathcal{D}_t$. This operation is repeated at each step to take advantage of the knowledge learned in previous steps and results in the creation of a hybrid dataset $\tilde{\mathcal{D}}$ composed by real samples $(\mathbf{X}_i,y_i)$ and samples with generated ground truth $(\mathbf{X}_i,\tilde{y}_i)$.

A key decision is how to generate the pseudo-labels $\tilde{y}$ for the $N_u$ unlabeled utterances. In this paper, we experiment with the use of hard pseudo-labels and soft pseudo-labels.
\paragraph{Hard pseudo-labels}
\label{sec:hard-pseudo}
In this approach hard pseudo-labels are directly obtained from network predictions. Let $\mathbf{p}_i$ be the probability outputs of our trained $h_\theta$ model for the utterance. Using the probability vector, the pseudo-label for the utterance $\mathbf{X}_i$ corresponds to $\tilde{y}_i = \mathrm{arg\; max}(\mathbf{p}_i)$. We select the subset of pseudo-labels which are less noisy to limit the confirmation bias, i.e. the overfitting of incorrect pseudo-labels predicted by the model. In particular, we select only the pseudo-labels corresponding to high-confidence predictions. Let $\mathbf{g} = \{g_b\}_{b=1}^B$ be a binary vector representing the selected pseudo-labels in a mini-batch $B$. This vector is obtained as follows:
\begin{equation}
    g_b = \mathbbm{1} \left[\mathrm{max}(p_b) \ge \tau \right],
\end{equation}
where $\tau \in (0,1)$ is a confidence threshold.

\paragraph{Soft pseudo-labels}
We investigate the use of soft pseudo-label inspired by \citep{arazo2020pseudo} since it has demonstrated in some cases to perform better than hard pseudo-labels~\citep{tanaka2018joint}.

Let $\mathbf{p}_i$ be the probability outputs of our trained $h_\theta$ model for the utterance $\mathbf{X}_i$. Two regularization terms are used to improve convergence. The first regularization term discourages the model to assign all samples to a single class by adding:
\begin{equation}
    R_A = \sum_{c=1}^C \mathbf{p}_c \mathrm{log}\left(\frac{\mathbf{p}_c}{\mathbf{\bar{h}}_c}\right),
\end{equation}
where $\mathbf{p}_c$ is the prior probability distribution for class $c$ assumed as a uniform distribution $\mathbf{p}_c = 1/C$ and $\mathbf{\bar{h}}_c$ denotes the mean softmax probability of the model for class $c$ across batch utterances.

The second regularization is the average per-sample entropy ($R_H$ stands for entropy regularization) that forces the probability distribution to peak on a single class:
\begin{equation}
    R_H = -\sum_{i=1}^B\sum_{c=1}^C h_\theta^c(\mathbf{X}_i)\mathrm{log}(h_\theta^c(\mathbf{X}_i)),
\end{equation}
where $h_\theta^c(\mathbf{X}_i)$ denotes the $c$ class value of the softmax output $h_\theta(\mathbf{X}_i)$ and it is estimated on a mini-batch $B$. The total loss is the following:
\begin{equation}
    \mathcal{L} = \mathcal{L}_{CE} + \lambda_AR_A + \lambda_HR_H,
\end{equation}
where $\lambda_A$ and $\lambda_H$ control the contribution of each regularization term.

To limit the confirmation bias problem, we exploit the mixup data augmentation technique proposed in \citep{zhang2018mixup}. It combines data augmentation with label smoothing to reduce the confidence of the model on its predictions. Mixup trains on convex combinations of sample pairs ($\mathbf{X}_p$ and $\mathbf{X}_q$) and corresponding labels ($y_p$ and $y_q$):
\begin{align}
    x = \alpha \mathbf{X}_p + (1-\alpha) \mathbf{X}_q,\\
    y = \alpha y_p + (1-\alpha) y_q,
\end{align}
where $\alpha$ is randomly sampled in the range $(0,1)$.
\subsubsection{Utterance rebalancing}
\label{sec:utt-rebal}
As stated in Section \ref{sec:problem-formulation}, the number of labeled utterances for the new language $N_k$ is much lower than the number of labeled utterances for the source language $N_s$. The use of pseudo-labeling to adapt the model to the new language can only partially reduce the imbalance between source and new language corpus.
The imbalanced ratio $\gamma_l$ between the number of source utterances, $N_s$, and the number of labeled utterances for the new language, $N_k$, is defined as $N_k / N_s$ and a $\gamma_l$ far from 1 indicates more severe utterance imbalance.

To tackle the utterance imbalance and get $\gamma_l = 1$ we exploit random oversampling. Specifically, the utterances of the new language are randomly replicated to match the number of utterances of the source language.

The rebalancing algorithm is run at the end of each pseudo-labeling iteration during the adaptation procedure. Furthermore, it is performed at the beginning of the training procedure if the number of labeled utterances for the new language is non-zero.
\subsubsection{Training procedure}
The proposed model $f_\theta(\mathbf{X})$ is trained end-to-end for $E$ epochs using a set of $N$ training utterances belonging to the joint domain $\mathcal{D} = \{\mathcal{D}_s \cup \mathcal{K}_t\}$. $\mathcal{D}$ contains a set of $N_s + N_k$ labeled utterances $\{\mathcal{D}_s, \mathcal{K}_t\}$ coming from both the source and the target language. Every $\textit{W}$ epochs, the model adaptation procedure is performed, in which supervised learning is accompanied by pseudo-labeling on the set $\mathcal{U}_t$ of unlabeled utterances for the new language. The set $\mathcal{D}$ is then expanded with the generated pseudo-labels $\mathcal{K}'_t\}$. The complete training procedure is presented in Algorithm \ref{alg:data-rebalance}.

The parameters $\theta$ of the model are optimized using categorical cross-entropy:
\begin{equation}
    \mathcal{L}_{CE} = - \sum_{i=1}^By_i\mathrm{log}(f_\theta(\mathbf{X}_i)),
\end{equation}
where $f_\theta(x)$ are the softmax probabilities predicted by the model, $\mathrm{log(\cdot)}$ is applied element-wise and $y_i$ can be a real or pseudo label, and $B$ is the number of batch utterances. 

\ADD{To mitigate the risk of overfitting, early stopping is implemented by selecting the model weights from the epoch that achieves the best performance on the validation set for both the source and new languages.}
\begin{algorithm}
\caption{Our training procedure.}
\label{alg:data-rebalance}
\begin{algorithmic}[1]
\STATE \textbf{Input:} Total training epochs $E$, interval of epochs for pseudo-labeling $W$, source language corpus $\mathcal{D}_s$, labeled utterances for the new language $\mathcal{K}_t$, unlabeled utterances for the new language $\mathcal{U}_t$.
\STATE Initialize the model $f_\theta$.
\STATE Initialize labeled corpus $\mathcal{D} = \mathcal{D}_s \cup \mathcal{K}_t$.
\STATE \textbf{for} e=1 to \textit{E} \textbf{do}
\STATE \hspace{0.5cm} Train and update $f_\theta$ on $\mathcal{D}$.

\STATE \hspace{0.5cm} \textbf{if} \textbf{mod}(e, \textit{W}) = 0 \textbf{then}
\STATE \hspace{1cm} Generate pseudo-labels for $\mathcal{U}_t$ using $f_\theta$.
\STATE \hspace{1cm} Form $\mathcal{K}'_t$ by applying pseudo-label policy on $\mathcal{U}_t$.
\STATE \hspace{1cm} Expand labeled set by $\mathcal{D} = \mathcal{D} \cup \mathcal{K}'_t$.
\STATE \hspace{0.5cm} \textbf{end if}
\STATE \textbf{end for}

\STATE \textbf{Return:} $f_\theta$
\end{algorithmic}
\end{algorithm}

\section{Experiments}
\label{sec:experiments}
In this section the datasets considered for experiments and the experimental setup are presented.
\subsection{Datasets}
A summary of the datasets used for our experiments is presented in Table  \ref{tab:emotion-dbs}. We consider seven speech emotion classification datasets in five languages: three in English (RAVDESS,  SAVEE and TESS), one in French (CaFE), German (EmoDB), Italian (EMOVO), and Persian (ShEMO). In each dataset, speech samples have three attributes: audio data (i.e., the raw waveform, in mono), speaker identifier, and emotion label (e.g., angry, happy, sad). The datasets comprise scripted and acted utterances and vary in size (i.e., number of utterances), number of speakers, sample rate, and number of classes. All of them comprise utterances in which the speaker acts a specific emotion.
\begin{table}
    \centering
    \caption{List of considered datasets for speaker emotion recognition.}
    \label{tab:emotion-dbs}
    \resizebox{\linewidth}{!}{\begin{tabular}{lcccccccc}
    \toprule
    Name & Spkrs & Emot. & SR (Hz) & Utter. & Lang. & \makecell{Avg. \\ dur. (s)} & \makecell{Tot. \\ dur. (h)} \\ \midrule
    CaFE \citep{gournay2018canadian} & 12 & 7 & 48,000 & 864 & French & 4.5 & 1.1 \\
    EMO-DB \citep{burkhardt2005database} & 10 & 7 & 16,000 & 535 & German & 2.8 & 0.4 \\
    EMOVO \citep{costantini2014emovo} & 6 & 7 & 48,000 & 588 & Italian & 3.1 & 0.5 \\
    RAVDESS \citep{livingstone2018ryerson} & 24 & 8 & 48,000 & 1,440 & English & 3.7 & 1.5 \\
    SAVEE \citep{wang2010machine} & 4 & 7 & 44,100 & 480 & English & 3.8 & 0.5 \\
    ShEMO \citep{nezami2019shemo} & 87 & 6 & 44,100 & 3,000 & Persian & 4.0 & 3.3 \\
    TESS \citep{SP2/E8H2MF_2020} & 2 & 7 & 24,414 & 2,800 & English & 2.0 & 1.6 \\
    \bottomrule
    \end{tabular}}
\end{table}

The considered datasets share the same five primary emotions, which are anger (A), fear (F), happiness (H), neutral (N), and sadness (S). Following \citep{tamulevivcius2020study}, in this study we consider only utterances annotated with one of the previous five emotions and discard the remaining utterances. Thus, the number of emotion categories is $C = 5$. TESS, SAVEE, and RAVDESS datasets are merged to obtain a large English language dataset. A summary of the distributions of utterances by emotion for each language is shown in the Table \ref{tab:balanced_samples}.
\begin{table}
    \centering
    \caption{Utterance distribution of the selected datasets across emotion classes.}
    \label{tab:balanced_samples}
    \begin{tabular}{lccccccc}
    \toprule
    & A & F & H & N & S &	Total\\
    \midrule
    \makecell[l]{English (RAVDESS, \\SAVEE, TESS)} & 652 & 652 & 652 &  652 & 616 & 3224 \\
    French (CaFE) &                               144 & 144 & 144 &  144 &  72 & 648 \\
    German (EMO-DB) &	                          127 &  69 &  71 &   79 &  62 & 408 \\
    Italian (EMOVO) &                              84 &  84 &  84 &   84 &  84 & 420 \\
    Persian (ShEMO) &                            1059 &  38 & 201 & 1028 & 449 & 2775 \\
    \bottomrule
    \end{tabular}
\end{table}

\subsection{Experimental setup}
Each dataset was split into training, validation, and testing sets to respectively train, optimize and evaluate task-specific emotion speech classifiers. Following SERAB \citep{scheidwasser2022serab}, each dataset is split into 60\% training, 20\% validation, and 20\% testing sets. Each data partition is speaker-independent, i.e., the sets of speakers included in each part are mutually disjoint.

Importantly, the utterances of the various datasets have different sampling rates, for this reason they were all resampled to 16kHz before any processing. During training a sequence of 2 seconds is randomly sampled from the whole utterance for augmentation, while the whole utterance is used at testing time. Voice Activity Detection (VAD) is not used in training and testing. The linear layer for emotion speech classification is randomly initialized.

All experiments are run three times with different random seeds, and the unweighted accuracy is chosen as our evaluation criterion.

\subsubsection{Hyperparameters}
In our experiments, we train the model for a total of 100 epochs using the Adam optimizer with an initial learning rate equal to $1 \times 10^{-3}$ which decays by a factor of 0.95 every 10 epochs, a batch of 100 utterances, and exponential decay rates $\beta_1$ and $\beta_2$ equal to 0.9 and 0.999. The pseudo-labeling procedure is executed every 30 epochs, i.e. $\textit{W}=30$. We experiment with different values of $\tau$ (see Sec. \ref{sec:tau-ablation} for a study of this hyperparameter) which lead to the choice of $\tau=0.50$, but do not attempt a careful adjustment of the regularization weights $\lambda_A$ and $\lambda_H $ and simply set them to 0.8 and 0.4 as done in \citep{tanaka2018joint}.
\section{Results}
\label{sec:results}
In this section the results achieved for different configurations are described. In all the experiments we consider English as the first language while the other languages were chosen one at a time as the second language.
%
\subsection{Cross-lingual results}
\label{x-ling-res}
In this section the performance obtainable by our method on a totally unknown new language is measured. These results give an idea of the worst accuracy, defined as the lower bound, achievable for cross-lingual SER. To this end, experiments using only the training set of the source language (i.e. English) and testing on the new language data are performed. Table \ref{tab:lower-result} reports accuracy results achieved by the three utterance encoders for the four new languages. As it is possible to see, HuBERT achieves the best accuracy for all the languages, while the EmotionCNN obtains the worst performance. Regarding HuBERT, the highest accuracy equal to 78.87\% is achieved for the German language while the lowest accuracy of 33.24\% is obtained for Persian. The same gap is registered for all utterance encoders and depends on the fact that as stated by several linguistic distances \citep{chiswick2005linguistic, petroni2008language,gamallo2017language} Persian belongs to a linguistic strain very different from the English one. Therefore, it is conceivable that learning on utterance in Persian will produce an important gain in performance.
\begin{table}
    \centering
    \caption{SER accuracy on new languages by training only on the source language (English). Results are reported for three different versions of the proposed model in which a different utterance encoder is exploited. Best result for each utterance encoder is in \textbf{bold}.}
    \label{tab:lower-result}
    \begin{tabular}{lccc}
    \toprule
         & \multicolumn{3}{c}{Accuracy (\%)} \\
         & EmotionCNN & BYOL-S & HuBERT \\ \midrule
        French & 29.63 & 54.32 & \textbf{58.02} \\
        German & 29.58 & 60.56 & \textbf{78.87} \\
        Italian & 26.43 & 49.28 & \textbf{52.14} \\
        Persian & 15.15 & 18.28 & \textbf{33.24} \\ \bottomrule
    \end{tabular}
\end{table}
\subsection{Multi-lingual results}
\label{multi-ling-res}
In this section, multi-lingual SER experiments are performed following a supervised training that uses all data of the source and the new language. The results obtained give the upper bound for cross-lingual SER, i.e. the best accuracy obtainable having all the labels of the new language available. Since we want to evaluate whether the SER classifier generalizes on the new language but also if it preserves the performance on the source language. Table \ref{tab:upper-result} shows the accuracy on the source language (i.e. first column of the table) and the accuracy on the new language (i.e. second column of the table) achieved by the three utterance encoders.

From the results it is possible to make various considerations. First, as expected, there is an increase in performance on the new language by using the training data of both the source language and the new one. This result is particularly evident for Persian, where the performance increases of about 50\% for all the utterance encoders. Second, EmotionCNN achieves significantly lower performance than the other two encoders, i.e. about 20\% less than BYOL-S and 30\% less than HuBERT. Third, the performance on English, which is the source language, does not degrade by adding the training data of the new language. For HuBERT, which confirms itself as the best encoder, starting from the 81.50\% of accuracy obtained by training only on English, we obtain a loss of about 3\% for both French and Persian. On the other hand, for Italian and German, the variation in accuracy is less than 1\%. 
\begin{table}
    \caption{SER accuracy on new languages by training on the combination of source and new language training sets. Results are reported for three different versions of the proposed model in which a different utterance encoder is exploited. Best result for each utterance encoder is in \textbf{bold}.}
    \label{tab:upper-result}
    \centering
    \resizebox{\linewidth}{!}{\begin{tabular}{lccc|ccc}
    \toprule
         & \multicolumn{3}{c|}{Accuracy on the source language (\%)} & \multicolumn{3}{c}{Accuracy on the new language (\%)} \\
        & EmotionCNN & BYOL-S & HuBERT & EmotionCNN & BYOL-S & HuBERT \\ \midrule
        English & 58.50 & 79.13 & \textbf{81.50} & -- & -- & -- \\ \midrule
        English \& French & 57.32 & 60.08 & \textbf{78.19} & 39.63 & 58.02 & \textbf{77.78} \\
        English \& German & 52.68 & 75.04 & \textbf{82.44} & 54.93 & 73.24 & \textbf{94.37} \\
        English \& Italian & 53.31 & 73.78 & \textbf{81.02} & 32.86 & 59.28 & \textbf{74.29} \\
        English \& Persian & 51.57 & 71.26 & \textbf{78.50} & 62.02 & 86.43 & \textbf{92.24} \\ \bottomrule
    \end{tabular}}
\end{table}
\subsection{Pseudo-labeling results}
\begin{figure*}
    \centering
    \begin{subfigure}{.85\linewidth}
        \centering
        \includegraphics[width=\textwidth]{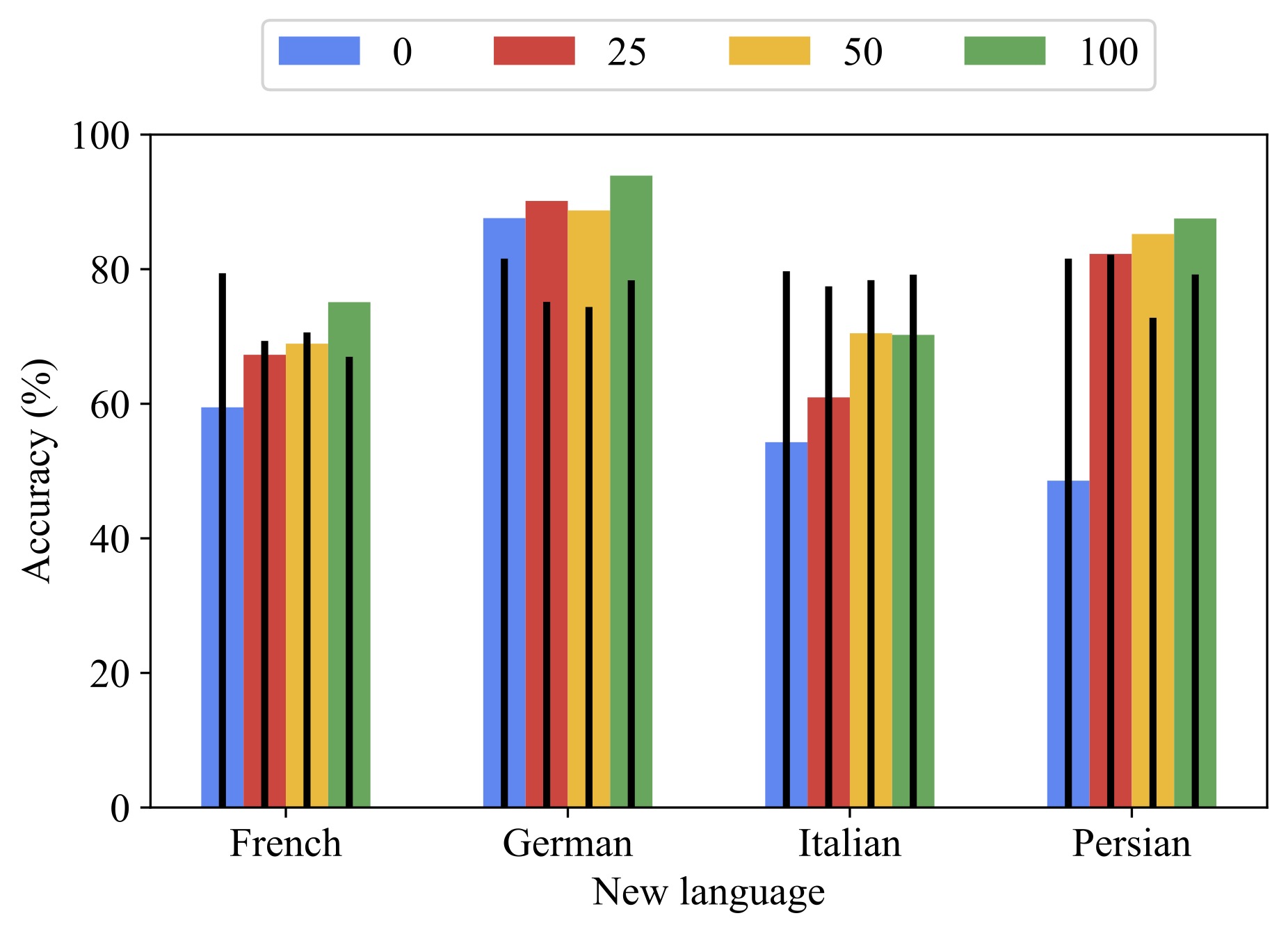}
        \caption{Results for hard pseudo-labels}
        \label{fig:result-hard-labeling}
    \end{subfigure}\\
    \begin{subfigure}{.85\linewidth}
        \centering
        \includegraphics[width=\textwidth]{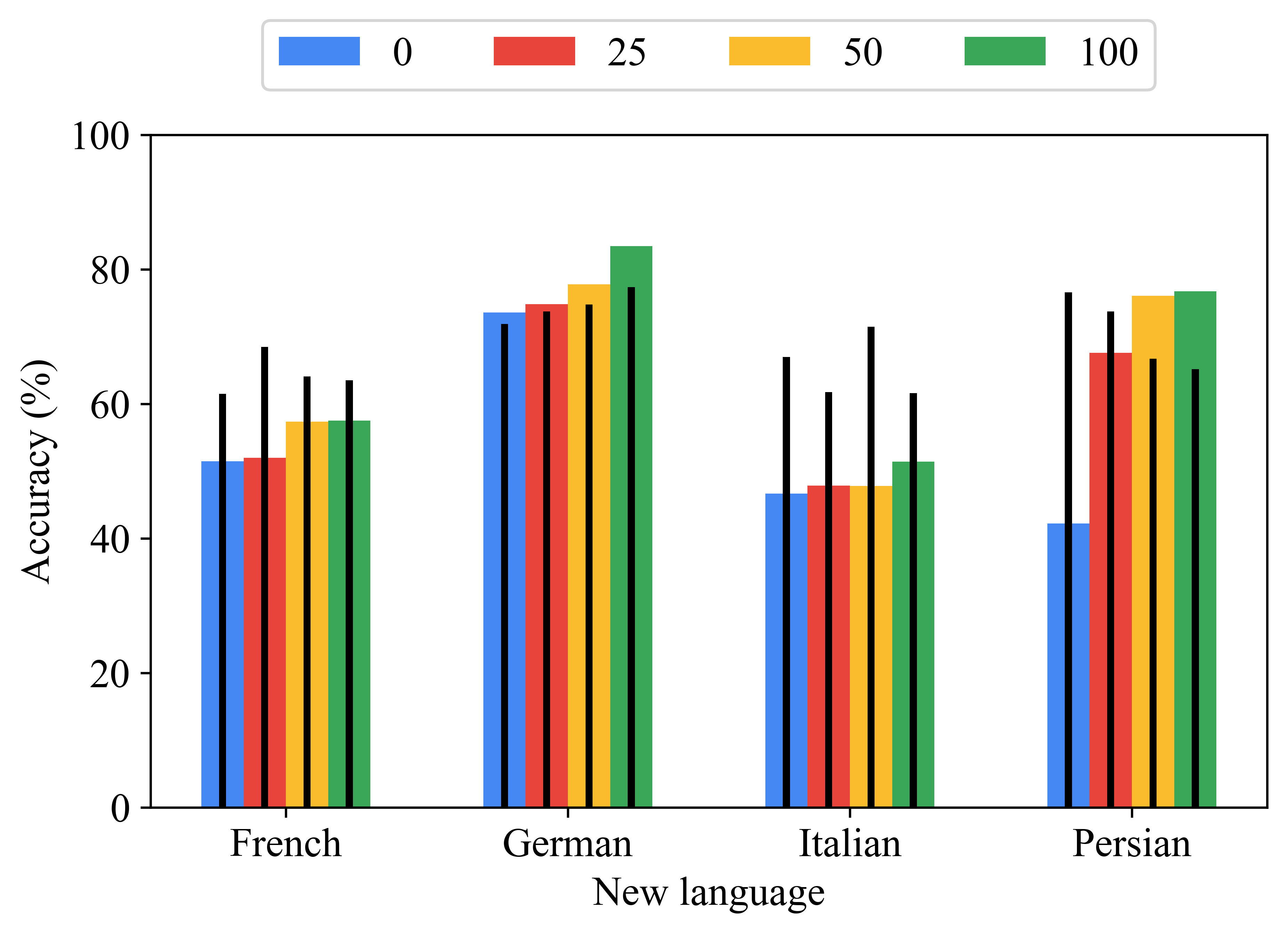}
        \caption{Results for soft pseudo-labels}
        \label{fig:result-soft-labeling}
    \end{subfigure}
    \caption{SER accuracy by varying the number of labeled utterances for the new language and using (a) hard pseudo-labels and (b) soft pseudo-labels for the unlabeled utterances. Black bars indicate the accuracy on the source language (i.e. English).}
    \label{fig:result-comparison}
\end{figure*}

In this subsection, the results of using the proposed method with HuBERT for cross-lingual SER are presented. Performance is reported while changing the number of labeled utterances for the new language. The numbers of labeled utterances considered for the new language are 0, 25, 50, and 100, while all the utterances for the source language are labeled (i.e. 1682 utterances). From the previous numbers we obtain the imbalanced ratios, $\gamma_l$, equal to 0, 0.01, 0.03 and 0.06. Figure \ref{fig:result-comparison} shows the performance achieved by using hard pseudo-labels (see Fig. \ref{fig:result-hard-labeling}) and soft pseudo-labels (see Fig. \ref{fig:result-soft-labeling}). Each colored bar represents the accuracy of a given pseudo-label approach with a given number of available labels for the new language. Black narrow bars represent the accuracy on the source language.

Several considerations can be made. First, overall hard pseudo-labels obtain better performance than soft pseudo-labels. Second, soft pseudo-labels not only result in worse performance than hard pseudo-labels, but also cause significant performance degradation on the source language. This behavior can be due to the effect of mixup augmentation which results in too noisy pseudo-labels that do not allow the model to converge properly. Third, for both pseudo-label approaches and for all languages, having more available labels for the new language allows to achieve higher accuracy. The pseudo-label approaches cannot reach the upper bound in any language but in any case they manage to improve the performance of the lower bound (for the Persian language it is even possible to have an increase of about 60\%).

\begin{figure}
    \centering
    \setlength{\tabcolsep}{1pt}
    \ADD{
    \begin{tabular}{cccc}
         French & German & Italian & Persian \\
         \includegraphics[height=.33\linewidth]{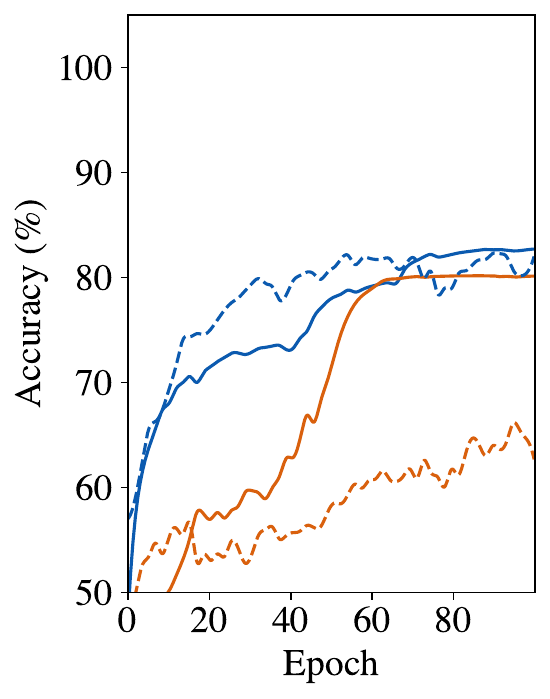} & \includegraphics[height=.33\linewidth]{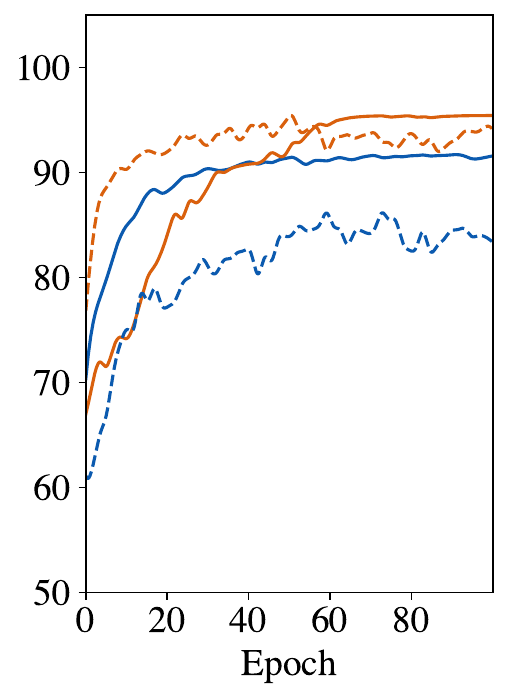} & \includegraphics[height=.33\linewidth]{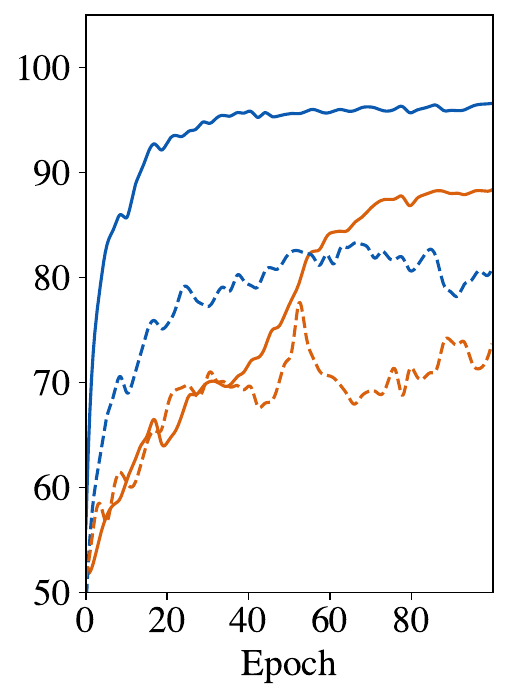} & \includegraphics[height=.33\linewidth]{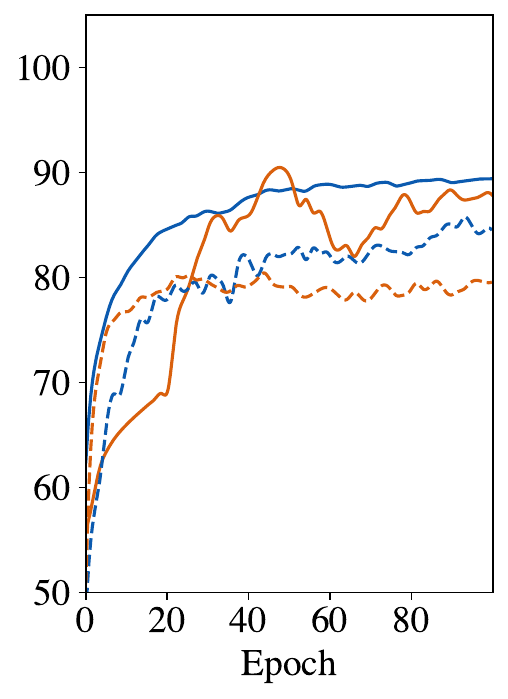} \\
         \multicolumn{4}{c}{\includegraphics[width=.45\linewidth]{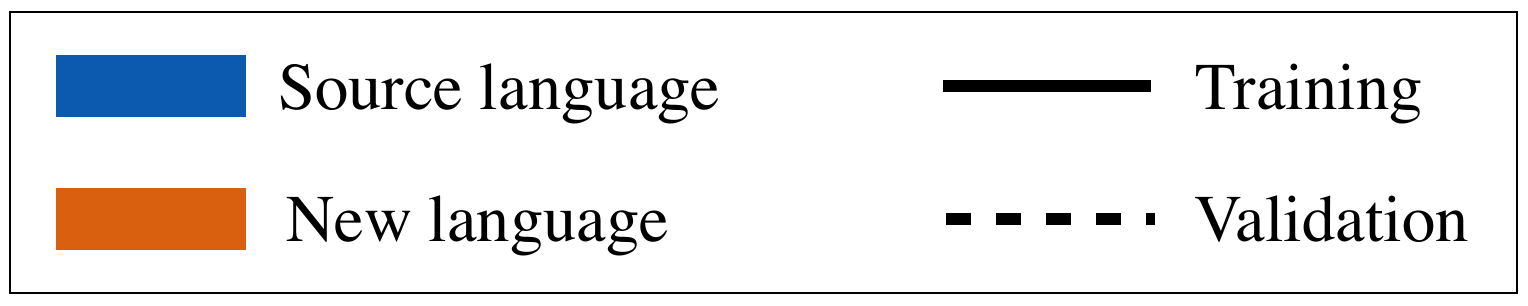}}
    \end{tabular}}
    \caption{\ADD{Accuracy curves for training and validation sets over the training of HuBERT with 100 labeled utterances for the new language. Each plot reports the accuracy on both the source language, i.e. English, and on each of the languages considered \textit{(better see in color and magnified)}.}}
    \label{fig:training-accuracy}
\end{figure}

\ADD{Figure \ref{fig:training-accuracy} illustrates the accuracy trends of the SER model with HuBERT and 100 labeled utterances across training epochs, depicting the performance of the training and validation sets. Each plot showcases accuracy curves for both the source and new languages. Despite having disjoint sets of speakers, the accuracy remains relatively consistent across the train and validation sets. However, the French and Italian languages, identified as the weakest performers, exhibit a tendency to overfit the model.}
\subsection{Discussion}
\begin{figure}
    \centering
    \includegraphics[width=.85\linewidth]{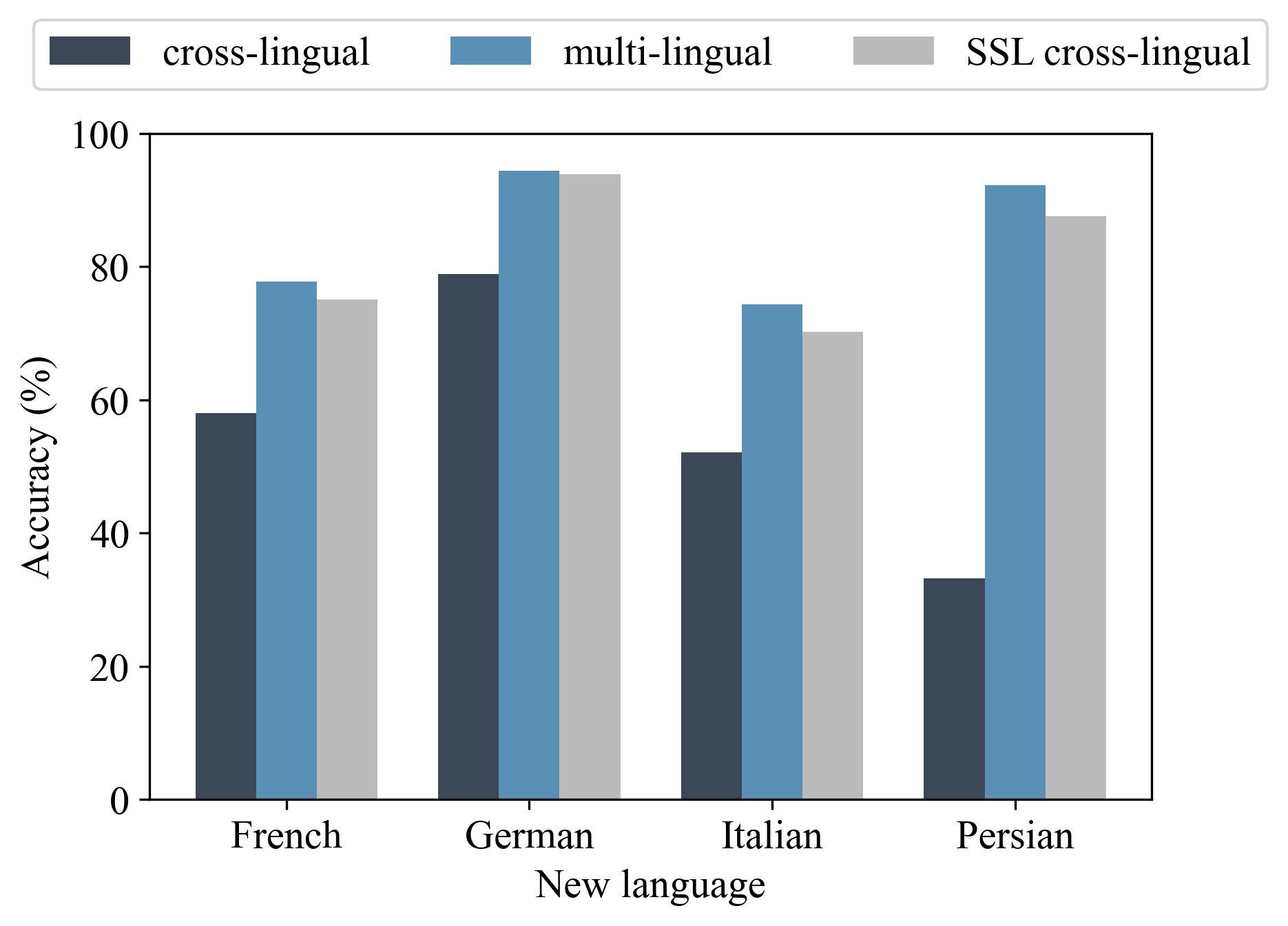}
    \caption{Comparison of the results on new languages between the cross-lingual, multi-lingual and the best SSL (hard pseudo-labels considering 100 labeled utterances).
    }
    \label{fig:result-summary}
\end{figure}

Figure \ref{fig:result-summary} summarizes the results achieved by our best method with HuBERT in the different configurations evaluated for the recognition of emotions on new languages. The aim is to highlight the gap between training the method with all the data labeled or with the use of pseudo-labeled samples. In the chart we compare the performance for (i) the cross-lingual experiment (results collected from the Table \ref{tab:lower-result}), (ii) the multi-lingual experiment (accuracy reported in the Table \ref{tab:upper-result}), and (iii) the SSL cross-lingual experiment based on the use of hard pseudo-labels having 100 labeled utterances for the new language (see Fig. \ref{fig:result-hard-labeling}). As expected, multi-lingual SER with all the utterances labeled results in a noticeable increase in performance compared to the cross-lingual SER. This increase is particularly significant for Persian (+60\%), a language that has very different linguistic traits from those of English and for this reason the adaptation of the model is very important. Cross-lingual SSL based on hard pseudo-labels improves the performance of the cross-lingual configuration for all the considered languages (the increase is 50\% for the Persian language).

An analysis of the feature space before and after the adaptation to the new language is also provided. The analysis aims to verify whether the pseudo-label approach can effectively reduce the variations between different languages while retaining information related to emotions. To illustrate this, we use Principal Component Analysis (PCA) to project the learned feature representation, i.e., the output of the utterance encoder into 3D space. Furthermore, the silhouette score \citep{rousseeuw1987silhouettes} is exploited to estimate the ability of the learned representation to discriminate emotions independently of language. The silhouette's best score value is 1 and the worst value is -1. Values close to 0 indicate overlapping clusters.

\begin{figure}
    \centering
    \setlength{\tabcolsep}{0pt}
    \begin{tabular}{cc|c|c|c}
        & French & German & Italian & Persian \\
        \rotatebox{90}{After training} & \includegraphics[width=.24\linewidth]{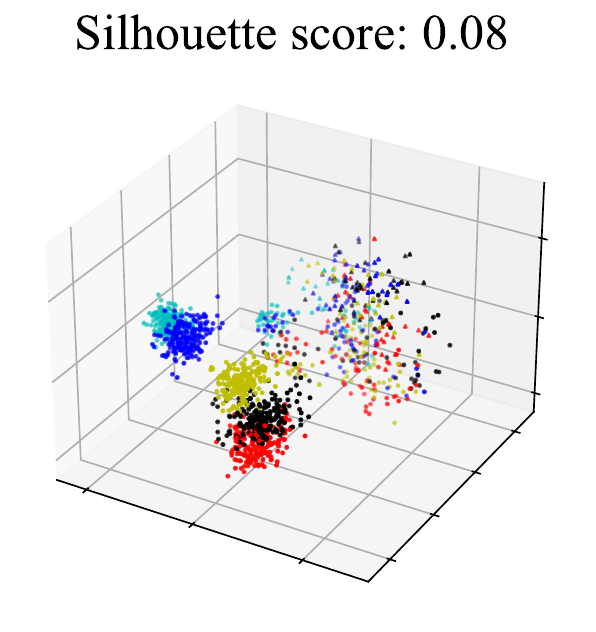} & \includegraphics[width=.24\linewidth]{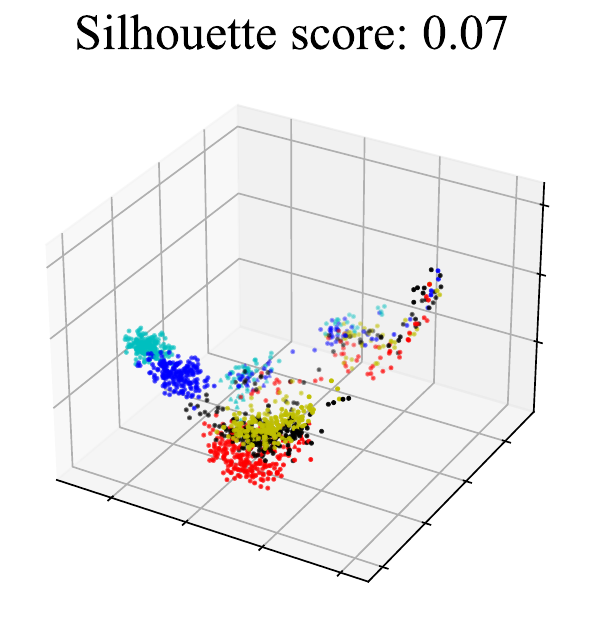} & \includegraphics[width=.24\linewidth]{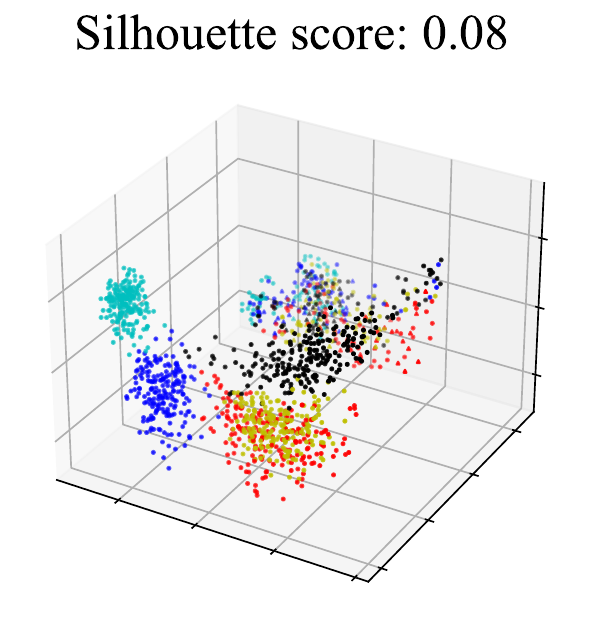} & \includegraphics[width=.24\linewidth]{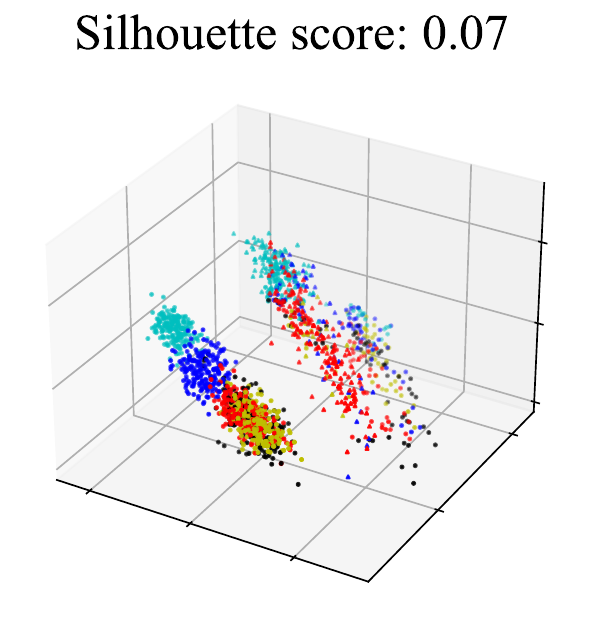} \\
        & $\Downarrow$ & $\Downarrow$ & $\Downarrow$ & $\Downarrow$\\
        \rotatebox{90}{After adaptation} & \includegraphics[width=.24\linewidth]{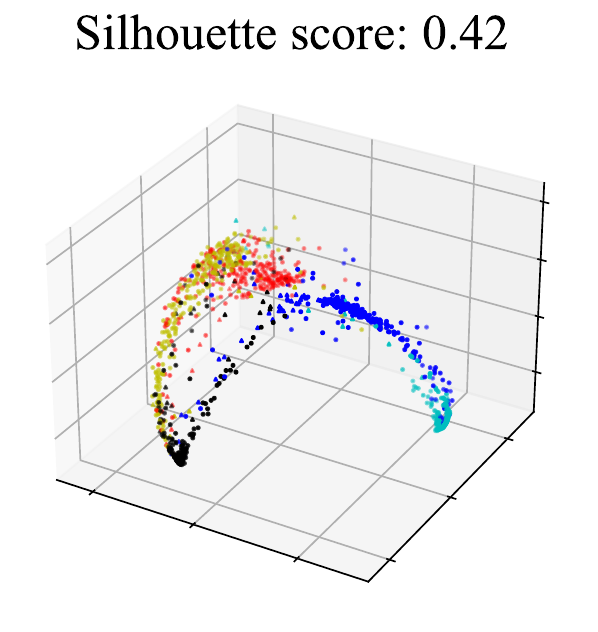} & \includegraphics[width=.24\linewidth]{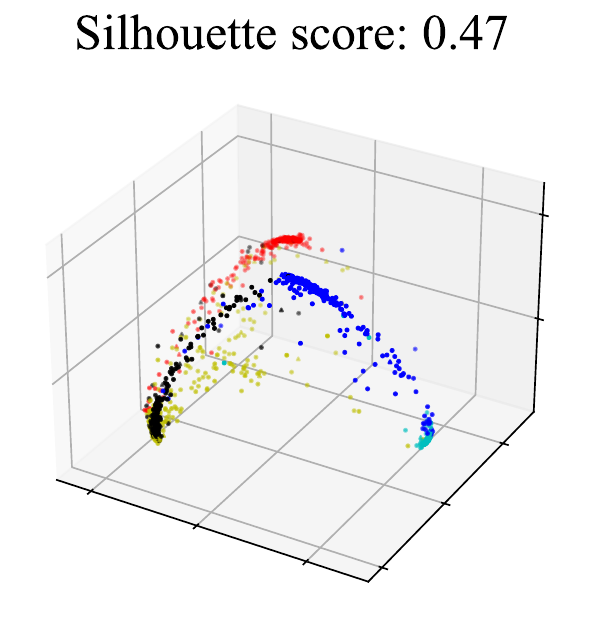} & \includegraphics[width=.24\linewidth]{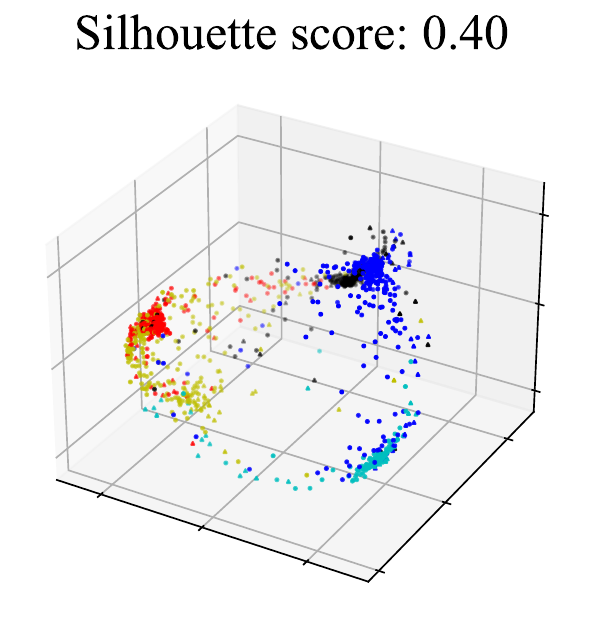} & \includegraphics[width=.24\linewidth]{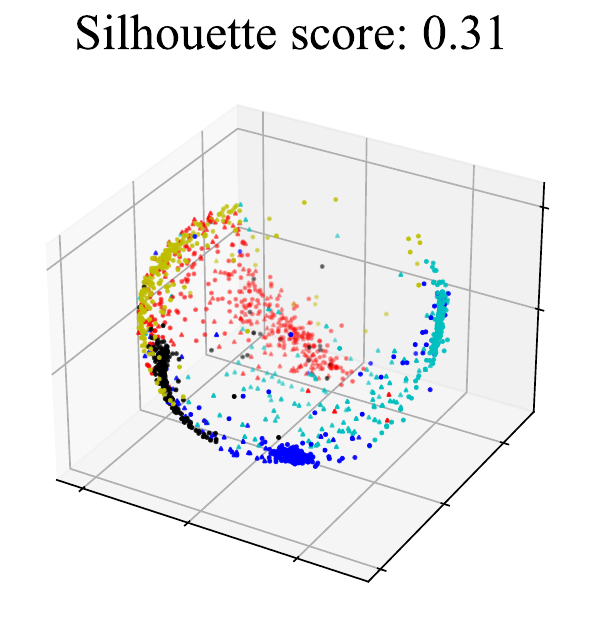} \\
        \multicolumn{5}{c}{\includegraphics[width=.9\linewidth]{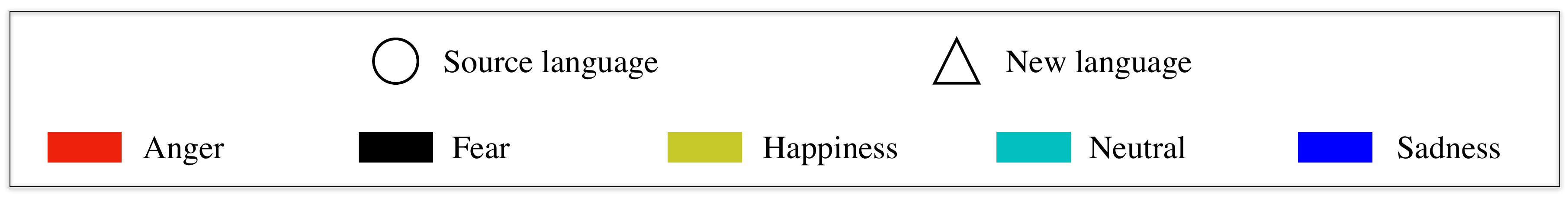}}
    \end{tabular}
    \caption{PCA plot of the learned feature representation with emotion and language labels after training on the English language only (first row) and after adaptation on the new language (second row) \textit{(better see in color and magnified)}.}
    \label{fig:pca}
\end{figure}

The first row of Figure \ref{fig:pca} displays the test utterances for both English and the new language encoded using HuBERT trained solely on English. Conversely, the second row shows the test utterances for both English and the new language encoded using HuBERT adapted on the new language via hard pseudo-labeling, and without any labeled utterances for the new language. The results indicate that the representation learned solely on English is unable to capture emotions for the new language, resulting in a very low silhouette score of approximately 0.07. In contrast, the adaptation procedure produces well-defined clusters based on emotion, independent of language, as evidenced by an average increase in the silhouette index of 0.36.
\subsection{Comparison with state-of-the-art}
In this section, we compare the performance of the proposed method with four recent state-of-the-art methods, AL~\citep{kim2017towards}, DANN~\citep{abdelwahab2018domain}, FLUDA~\citep{ahn2021cross}, and NNPM~\citep{li2021unsupervised}. The above methods are trained for discriminating 4 emotion categories namely anger, happiness, neutral, and sadness. Implementations of the methods are not publicly available so we first report their performance on the German language from the original documents or reimplementations. For a more in-depth comparative analysis we reimplemented the considered methods in order to be able to estimate the performance also on the other three languages, namely French, Italian and Persian. We only report the results obtained for DANN and NNPM, for which the performance are in line with those declared. For a fair comparison with previous methods, we exclude from the all corpus the utterances labeled with emotion of fear and train the methods.

Our method is retrained without using labeled utterances of the new language and taking advantage of hard pseudo-labels. Table \ref{tab:comparison} shows the comparison between two versions of the proposed method, i.e. with the backbone based on BYOL-S and HuBERT, and other methods. The results achieved for our implementations are respectively ``DANN (our reimplementation)'' and ``NNPM (our reimplementation)''. Results show that both versions of the proposed method perform better than recent state-of-the-art methods. More specifically, our HuBERT-based method outperforms the second best method, which is also based on pseudo-labeling, namely NNPM, with an improvement in relative accuracy of 30\%. We get better performance than the multi-task learning method, i.e. AL, with 57\% better accuracy, and the unsupervised cross-corpus SER model based on few-shot learning (i.e., FLUDA) of an increase of 64\%.

This high gain in performance with respect to previous methods is due to three main aspects. First, the state-of-the-art methods consist of very simple architectures compared to that of HuBERT. Second, the model training procedure is profoundly different between the previous methods and that used for HuBERT. While the purpose of previous methods is to directly learn a specialized mapping of a speech signal into an emotion category, HuBERT and BYOL-S are trained to learn a general-purpose and robust representation of a speech signal. This last aspect allows the obtained representation to be more effective for the different tasks, including the recognition of emotions. Ultimately, the difference between HuBERT and BYOL-S is due to both the architectural aspect of the model and the cardinality of the dataset used for the pre-training. In fact, HuBERT is trained on a much larger and challenging dataset than the one used for BYOL-S.
\begin{table}
    \centering
    \caption{Comparison with other state-of-the-art methods on new languages. Best result for each language is in \textbf{bold}.}
    \label{tab:comparison}
    \begin{tabular}{lcccc}
    \toprule
        Method & French & German & Italian & Persian \\ \midrule
        DANN \citep{ahn2021cross}       & -- & 28.5 & -- & -- \\
        DANN (our reimplementation)     & 31.5 & 32.6 & 25.9 & 30.7 \\
        FLUDA \citep{ahn2021cross}      & -- & 34.9 & -- & -- \\
        AL \citep{ahn2021cross}         & -- & 42.5 & -- & -- \\
        NNPM \citep{li2021unsupervised} & -- & 50.6 & -- & -- \\
        NNPM (our reimplementation) & 39.7 & 52. & 40.2 & 56.0 \\ \midrule
        Our (BYOL-S)                    & 52.8 & 66.7 & 48.4 & 68.4 \\
        Our (HuBERT)                    & \textbf{70.7} & \textbf{89.0} & \textbf{66.9} & \textbf{79.6} \\ \bottomrule
    \end{tabular}
\end{table}
\subsection{Ablation study}
This section presents an ablation study of the main design choices that led to the definition of the final method. The adaptability to the new language of CNN vs. Transformer based utterance encoders is evaluated. The effect of different values for the hard pseudo-labeling $\tau$ parameter is investigated. Finally, the impact of utterance rebalancing on the performance is estimated.
\subsubsection{Utterance encoder comparison}
\begin{figure}
    \centering
    \includegraphics[width=.85\linewidth]{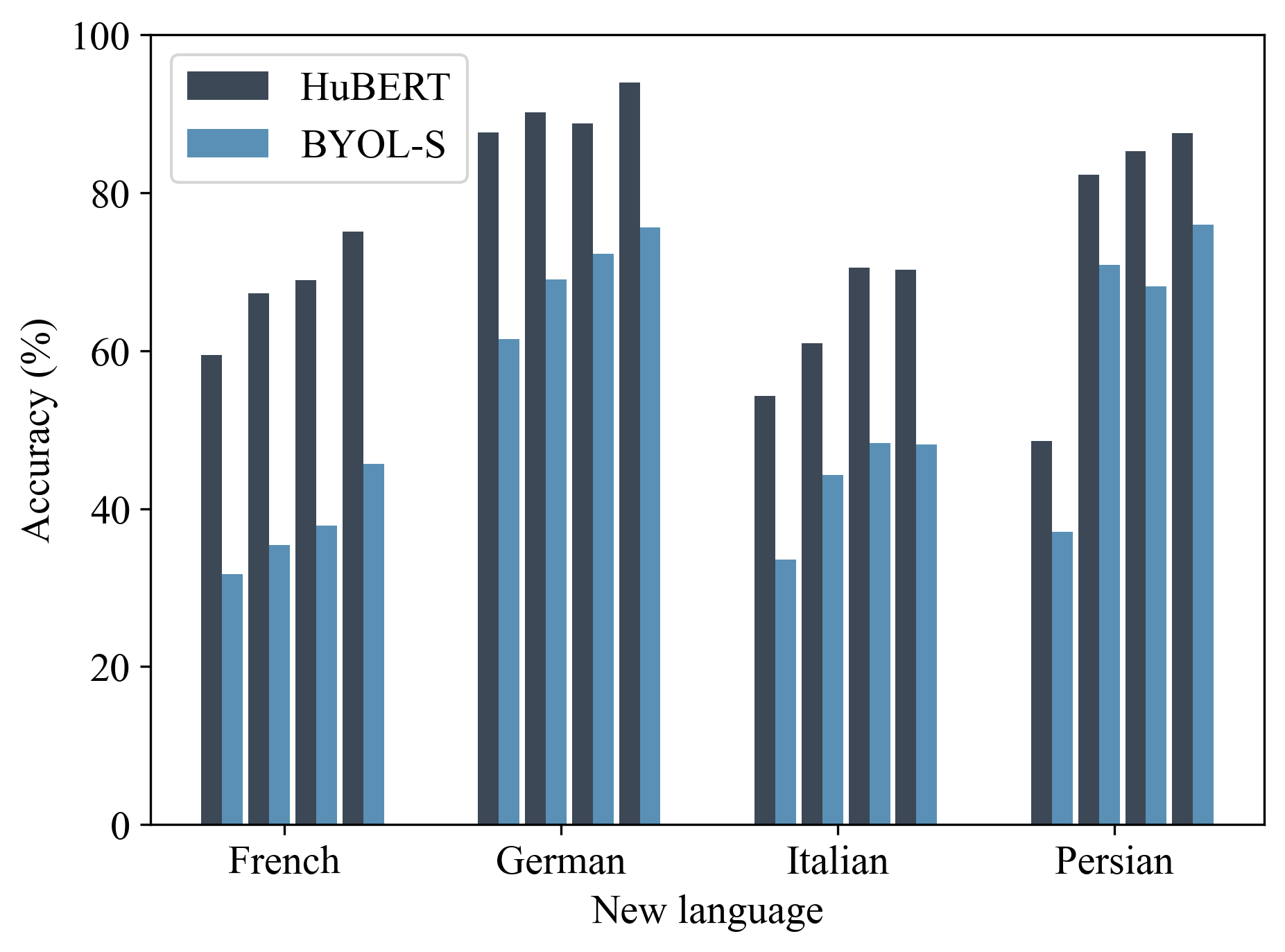}
    \caption{HuBERT vs. BYOL-S. Results for each new language using one or the other utterance embedding model and hard pseudo-labels. The number of utterances labeled for the new language is varied from 0 to 100.}
    \label{fig:result-ablation-hubert-vs-byols}
\end{figure}

Cross- and multi- lingual results demonstrate that HuBERT and BYOL-S provide more effective utterance encoding for emotion classification than EmotionCNN (see Sec. \ref{x-ling-res} and \ref{multi-ling-res}). In this section we perform the comparison for the different languages using hard pseudo-labels. The results for cross-lingual SER by varying the number of utterances labeled for the new language from 0 to 100 are shown in Figure \ref{fig:result-ablation-hubert-vs-byols}. As it is possible to see, HuBERT outperforms BYOL-S by a large gap (about +20\% accuracy). This gap might be motivated by several reasons. First, HuBERT is trained on a larger and more diverse speech corpus (i.e. Librispeech), with both spontaneous and anechoic scripted speech, while BYOL-S is  trained on a subset of AudioSet \citep{elbanna2022byols}. Second, HuBERT's transformer-based architecture coupled with direct encoding of the raw waveform provides a more robust and powerful representation of speech.
\subsubsection{Effect of $\tau$}
\label{sec:tau-ablation}
Among all the hyperparameters, the confidence threshold $\tau$ used for hard pseudo-labels (see Sec. \ref{sec:hard-pseudo}) is the one that needs to be carefully tuned. This subsection studies the effects of $\tau$ on our hard pseudo-labeling approach. Table \ref{fig:result-tau} reports the cross-lingual SER results by considering 100 labeled utterance for the new language and varying $\tau$. From the results it is possible to observe that the choice of the best $\tau$ does not generalize to all languages. However, for 0.50 the best performance is obtained in most cases and for this reason it was chosen.
\begin{table}
    \caption{SER accuracy obtained using 100 labeled utterances for the new language and varying $\tau$ values for hard pseudo-labeling. Best result for each language is in \textbf{bold}.}
    \label{fig:result-tau}
    \centering
    \begin{tabular}{lccc}
    \toprule
      &  $\tau = 0.50$ & $\tau = 0.65$ & $\tau = 0.85$ \\ \midrule
    English & 66.98 & 68.92 & \textbf{75.91} \\
    French  & \textbf{75.10} & 73.86 & 74.28 \\ \midrule
    English & \textbf{78.35} & 76.56 & 75.87 \\
    German  & 93.95 & \textbf{94.37} & 93.90 \\ \midrule
    English & 79.19 & \textbf{81.89} & 77.69 \\
    Italian & \textbf{74.24} & 68.81 & 73.10 \\ \midrule
    English & 79.21 & \textbf{81.47} & 79.13 \\
    Persian & \textbf{88.53} & 88.27 & 88.18 \\ \bottomrule
    \end{tabular}
\end{table}
\subsubsection{Effect of utterance rebalancing}
\label{sec:rebalance-ablation}
Here we quantitatively evaluate the contribution of the utterance rebalancing procedure on the performance of our method (see Sec. \ref{sec:utt-rebal} for details). Table \ref{tab:result-rebalance} shows the cross-lingual SER results of our method without utterance rebalance ``w/o rebalance'' and with utterance rebalance ``w rebalance''. As it is possible to see, the version of the method with utterance rebalancing outperforms the version without utterance rebalancing for all languages. The highest accuracy improvement corresponding to 40\% is registered for the Persian language. The lowest gaps of 8\% and 15\% between the two versions are obtained for French and Italian, respectively.
\begin{table}
    \centering
    \caption{SER accuracy obtained using 100 labeled utterances for the new language and hard pseudo-labeling with $\tau=0.5$. Performance without and with utterance rebalancing is compared. Best result for each language is in \textbf{bold}.}
    \label{tab:result-rebalance}
    \begin{tabular}{lcc}
    \toprule
         & w/o rebalance & w rebalance \\ \midrule
         English & 67.85 & \textbf{68.92} \\
         French & 65.64 & \textbf{73.86} \\ \midrule
         English & 74.13 & \textbf{76.56} \\
         German & 78.37 & \textbf{94.37} \\ \midrule
         English & 73.14 & \textbf{81.89} \\ 
         Italian & 53.64  & \textbf{68.81} \\ \midrule 
         English & 74.34 & \textbf{81.47} \\ 
         Persian & 47.42 & \textbf{88.27} \\ 
    \bottomrule
    \end{tabular}
\end{table}
%
%
\section{Conclusions}
\label{sec:conclusions}
In cross-lingual SER it is common to have many labeled utterances for the English language and a lower availability of labels for other languages. Based on this consideration, an SSL approach for cross-lingual speech emotion recognition is proposed.

The proposed method consists of a transformer able to classify an utterance into an emotion category. For SSL, we experimented with the use of hard and soft pseudo-labels for unlabeled utterances. The proposed method is evaluated using English as source language and four different languages (French, German, Italian and Persian) as new languages. It is revealed that the use of hard over soft pseudo-labels allows for better results on the new language at the expense of a drop in performance on the source language.

Experimental results show that the average accuracy has increased by 40\% in comparison with state-of-the-art methods.

The proposed method has some limitations, of course. First, the method assumes that the number of emotions in the source and the new language is the same. Nonetheless, in real-wold applications this constraint could be too stringent. To overcome this limitation, prototypes could be learned from representations for newly-introduced emotion categories in the target language, using a methodology similar to the one described in \citep{bucher2021handling}. Second, for the hard-pseudo labeling approach there is no handling of confirmation bias, i.e. overfitting to incorrect pseudo-labels predicted by the network. In fact, if the model makes several wrong unlabeled predictions, pseudo-labeling can act like a bad feedback loop and deteriorate performance. As future work we plan to handle the confirmation bias issue by averaging the predictions for different views of the unlabeled data as done in  \citep{berthelot2019mixmatch} or by using reinforcement learning \citep{latif2022survey}.

\bibliographystyle{model5-names}\biboptions{authoryear}

\end{document}